\documentclass[fleqn,11pt]{article}

\usepackage{latexsym,ifthen,epsfig,color}
\usepackage{amsmath,amssymb,amsthm}
\usepackage{float}

\newcommand{\join}{\text{\textcircled{{\footnotesize 1}}}}
\newcommand{\cojoin}{\text{\textcircled{{\footnotesize 0}}}}

\newcommand{\NP}{\ensuremath{\mathbb{NP}}}

\newtheorem{Claim}{Claim}[section]
\newtheorem{theorem}{Theorem}
\newtheorem{lemma}{Lemma}

\newtheorem{corollary}{Corollary}

\begin{document}

\author{
Andreas Brandst\"adt\\
\small Institut f\"ur Informatik, Universit\"at Rostock, D-18051 Rostock, Germany\\
\small \texttt{andreas.brandstaedt@uni-rostock.de}\\
\and
Raffaele Mosca\\
\small Dipartimento di Economia, Universit\'a degli Studi ``G. D'Annunzio'',
Pescara 65121, Italy\\
\small \texttt{r.mosca@unich.it}
}

\title{On Efficient Domination for Some Classes of $H$-Free Bipartite Graphs}

\maketitle

\begin{abstract}
A vertex set $D$ in a finite undirected graph $G$ is an {\em efficient dominating set} (\emph{e.d.s.}\ for short) of $G$ if every vertex of $G$ is dominated by exactly one vertex of $D$. The \emph{Efficient Domination} (ED) problem, which asks for the existence of an e.d.s.\ in $G$, is known to be \NP-complete even for very restricted $H$-free graph classes such as for $2P_3$-free chordal graphs while it is solvable in polynomial time for $P_6$-free graphs. Here we focus on bipartite graphs: We show that (weighted) ED can be solved in polynomial time for $H$-free bipartite graphs when $H$ is $P_7$ or $\ell P_4$ for fixed $\ell$, and similarly for $P_9$-free bipartite graphs with vertex degree at most 3, and  when $H$ is $S_{2,2,4}$. Moreover, we show that ED is \NP-complete for bipartite graphs with diameter at most 6.
\end{abstract}

\noindent{\small\textbf{Keywords}:
Weighted efficient domination;
$H$-free bipartite graphs;
\NP-completeness;
polynomial time algorithm;
clique-width.
}

\section{Introduction}\label{sec:intro}

Let $G=(V,E)$ be a finite undirected graph. A vertex $v$ {\em dominates} itself and its neighbors. A vertex subset $D \subseteq V$ is an {\em efficient dominating set} ({\em e.d.s.}\ for short) of $G$ if every vertex of $G$ is dominated by exactly one vertex in $D$; for any e.d.s.\ $D$ of $G$, $|D \cap N[v]| = 1$ for every $v \in V$ (where $N[v]$ denotes the closed neighborhood of $v$).
Note that not every graph has an e.d.s.; the {\sc Efficient Dominating Set} (ED) problem asks for the existence of an e.d.s.\ in a given graph~$G$.
The notion of efficient domination was introduced by Biggs \cite{Biggs1973} under the name {\em perfect code}.

\medskip

The Exact Cover Problem asks for a subset ${\cal F'}$ of a set family ${\cal F}$ over a ground set, say $V$, containing every vertex in $V$ exactly once, i.e., ${\cal F'}$ forms a partition of $V$. As shown by Karp \cite{Karp1972}, this problem is \NP-complete even for set families containing only $3$-element subsets of $V$ (see problem X3C [SP2] in \cite{GarJoh1979}).

Clearly, ED is the Exact Cover problem for the closed neighborhood hypergraph of $G$, i.e., if $D=\{d_1,\ldots,d_k\}$ is an e.d.s.\ of $G$ then $N[d_1] \cup \ldots \cup N[d_k]$ forms a partition of $V$ (we call it the {\em e.d.s.\ property}). In particular, the distance between any pair of distinct $D$-vertices $d_i,d_j$ is at least 3. 

\medskip

In \cite{BanBarSla1988,BanBarHosSla1996}, it was shown that the ED problem is \NP-complete. Moreover, Lu and Tang \cite{LuTan2002} showed that ED is \NP-complete for chordal bipartite graphs (i.e., hole-free bipartite graphs). Thus, for every $k \ge 3$, ED is \NP-complete for $C_{2k}$-free bipartite graphs.

Moreover, ED is \NP-complete for planar bipartite graphs \cite{LuTan2002} and even for planar bipartite graphs of maximum degree 3 \cite{BraMilNev2013} and girth at least $g$ for every fixed $g$ \cite{Nevri2014}. Thus, ED is \NP-complete for $K_{1,4}$-free bipartite graphs and for $C_4$-free bipartite graphs.

\medskip

In \cite{BraFicLeiMil2015}, it is shown that one can extend polynomial time algorithms for Efficient Domination to such algorithms for weighted Efficient Domination. Thus, from now on, we focus on the unweighted ED problem.

\medskip

\begin{figure}[ht]
  \begin{center}
   \epsfig{file=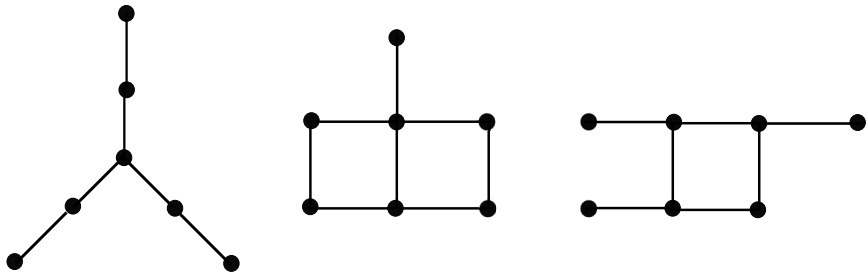}
   \caption{Forbidden induced subgraphs $H_1=S_{2,2,2},H_2,H_3$ for bipartite permutation graphs}
   \label{S222etc}
  \end{center}
\end{figure}

In \cite{BraLeiRau2012}, it is shown that ED is solvable in polynomial time for interval bigraphs, and
 convex bipartite graphs are a subclass of them (and of chordal bipartite graphs).
Moreover, Lu and Tang \cite{LuTan2002} showed that ED is solvable in linear time for bipartite permutation graphs (which is a subclass of convex bipartite graphs). It is well known (see e.g.\ \cite{BraLeSpi1999,Koehl1999}) that $G$ is a bipartite permutation graph if and only if $G$ is AT-free bipartite if and only if $G$ is ($H_1,H_2,H_3$,hole)-free bipartite (see Figure \ref{S222etc}).
Thus, while ED is \NP-complete for $(H_2,H_3)$-free bipartite graphs (since $H_2$ and $H_3$ contain $C_4$ and $H_2$ contains $K_{1,4}$), we will show that ED is solvable in polynomial time for $S_{2,2,2}$-free (and more generally, for $S_{2,2,4}$-free) bipartite graphs.

\medskip

For a set ${\cal F}$ of graphs, a graph $G$ is called {\em ${\cal F}$-free} if $G$ contains no induced subgraph isomorphic to a member of ${\cal F}$.
In particular, we say that $G$ is {\em $H$-free} if $G$ is $\{H\}$-free.
Let $H_1+H_2$ denote the disjoint union of graphs $H_1$ and $H_2$, and for $k \ge 2$, let $kH$ denote the disjoint union of $k$ copies of $H$.
For $i \ge 1$, let $P_i$ denote the chordless path with $i$ vertices, and let $K_i$ denote the complete graph with $i$ vertices (clearly, $P_2=K_2$).
For $i \ge 4$, let $C_i$ denote the chordless cycle with $i$ vertices.

For indices $i,j,k \ge 0$, let $S_{i,j,k}$ denote the graph with vertices $u,x_1,\ldots,x_i$, $y_1,\ldots,y_j$, $z_1,\ldots,z_k$ (and {\em center} $u$) such that the subgraph induced by $u,x_1,\ldots,x_i$ forms a $P_{i+1}$ ($u,x_1,\ldots,x_i$), the subgraph induced by $u,y_1,\ldots,y_j$ forms a $P_{j+1}$ ($u,y_1,\ldots,y_j$), and the subgraph induced by $u,z_1,\ldots,z_k$ forms a $P_{k+1}$ ($u,z_1,\ldots,z_k$), and there are no other edges in $S_{i,j,k}$. Thus, claw is $S_{1,1,1}$, chair is $S_{1,1,2}$, and $P_k$ is isomorphic to $S_{0,0,k-1}$.

\medskip

For a vertex $v \in V$, $N(v)=\{u \in V: uv \in E\}$ denotes its ({\em open}) {\em neighborhood}, and $N[v]=\{v\} \cup N(v)$ denotes its {\em closed neighborhood}.
The {\em non-neighborhood} of a vertex $v$ is $\overline{N}(v):=V \setminus N[v]$.
For $U \subseteq V$, $N(U):= \bigcup_{u \in U} N(u) \setminus U$ and $\overline{N}(U):=V \setminus(U \cup N(U))$.

We say that for a vertex set $X\subseteq V$, a vertex $v \notin X$ has a join (resp.,~co-join) to $X$ if $X\subseteq N(v)$ (resp., $X\subseteq \overline{N}(v)$).
Join (resp., co-join) of $v$ to $X$ is denoted by $v \join X$ (resp., $v \cojoin X$). Correspondingly, for vertex sets $X,Y \subseteq V$ with $X \cap Y = \emptyset$,
$X \join Y$ denotes $x \join Y$ for all $x \in X$ and $X \cojoin Y$ denotes $x \cojoin Y$ for all $x \in X$. A vertex $x \notin U$ {\em contacts $U$} if $x$ has a neighbor in $U$. For vertex sets $U,U'$ with $U \cap U' = \emptyset$, $U$ {\em contacts $U'$} if there is a vertex in $U$ contacting $U'$.

If $v \not\in X$ but $v$ has neither a join nor a co-join to $X$, then we say that $v$ {\it distinguishes} $X$.
A set $H$ of at least two vertices of a graph $G$ is called \emph{homogeneous} if $H \not= V(G)$ and every vertex outside $H$ is either adjacent to all vertices in $H$, or to no vertex in $H$.
Obviously, $H$ is homogeneous in $G$ if and only if $H$ is homogeneous in the complement graph $\overline{G}$.
A graph is {\em prime} if it contains no homogeneous set. In \cite{BraGia2014,BraGiaMil2018,BraMilNev2013}, it is shown that the ED problem can be reduced to prime graphs.

\medskip

It is well known that for a graph class with bounded clique-width, ED can be solved in polynomial time \cite{CouMakRot2000}. Thus we only consider ED on $H$-free bipartite graphs for which the clique-width is unbounded. In \cite{DabPau2016}, the clique-width of all classes of $H$-free bipartite graphs is classified.
For example, while ED is \NP-complete for claw-free graphs (even for ($K_{1,3},K_4-e$)-free perfect graphs \cite{LuTan1998}), the clique-width of claw-free bipartite graphs is bounded.

Based on Lozin's papers \cite{Lozin2002,LozRau2004,LozRau2006,LozVol2006} (with coauthors Rautenbach and Volz), the following dichotomy theorem was found by Dabrowski and Paulusma ($H \subseteq_i G$ indicates that $H$ is an induced subgraph of $G$):
\begin{theorem}[\cite{DabPau2016}]\label{cwdHfrbipgr}
The clique-width of $H$-free bipartite graphs is bounded if and only if one of the following cases appears:
\begin{enumerate}
\item[$(1)$] $H=sP_1$, $s \ge 1$;
\item[$(2)$] $H \subseteq_i K_{1,3} + 3P_1$;
\item[$(3)$] $H \subseteq_i K_{1,3} + P_2$;
\item[$(4)$] $H \subseteq_i S_{1,1,3} + P_1$;
\item[$(5)$] $H \subseteq_i S_{1,2,3}$.
\end{enumerate}
\end{theorem}

\medskip

Let $d_G(x,y)$ denote the distance of $x$ and $y$ in $G$. For graph $G=(V,E)$, its square $G^2$ has the same vertex set $V$, and two vertices $x,y \in V$,
$x \neq y$, are adjacent in $G^2$ if and only if $d_G(x,y) \le 2$. Let $N^2(v)=N(N(v))$, i.e.,  $N^2(v)$ is the subset of vertices which have distance 2 to $v$, and correspondingly, $N^2(U)$ for $U \subseteq V$ is the set of vertices which have distance 2 to at least one vertex of $U$.
The ED problem on $G$ can be reduced to Maximum Weight Independent Set (MWIS) on $G^2$ (see e.g.\ \cite{BraLeiRau2012,BraFicLeiMil2015,BraMilNev2013,Milan2012} and the survey in \cite{Brand2018}).

\section{ED in polynomial time for some $H$-free bipartite graphs}

\subsection{A general approach}\label{generalapproach}

A vertex $u \in V$ is {\em forced} if $u \in D$ for every e.d.s.\ $D$ of $G$; $u$ is {\em excluded} if $u \notin D$ for every e.d.s.\ $D$ of $G$.
By a forced vertex $u$, $G$ can be reduced to $G'$ as follows:
\begin{Claim}\label{forcedreduction}
If $u$ is forced then $G$ has an e.d.s.\ $D$ with $u \in D$ if and only if the reduced graph $G'=G \setminus N[u]$ has an e.d.s.\ $D'=D \setminus \{u\}$ such that all vertices in $N^2(u)$ are excluded in $G'$.
\end{Claim}

Analogously, if we assume that $v \in D$ for a vertex $v \in V$ then $u \in V$ is {\em $v$-forced} if $u \in D$ for every e.d.s.\ $D$ of $G$ with $v \in D$,
and $u$ is {\em $v$-excluded} if $u \notin D$ for every e.d.s.\ $D$ of $G$ with $v \in D$. For checking whether $G$ has an e.d.s.\ $D$ with $v \in D$, we can clearly reduce $G$ by forced vertices as well as by $v$-forced vertices when we assume that $v \in D$:

\begin{Claim}\label{vforcedreduction}
If we assume that $v \in D$ and $u$ is $v$-forced then $G$ has an e.d.s.\ $D$ with $v \in D$ if and only if the reduced graph $G'=G \setminus N[u]$ has an e.d.s.\ $D'=D \setminus \{u\}$ with $v \in D'$ such that all vertices in $N^2(u)$ are $v$-excluded in $G'$.
\end{Claim}

Similarly, for $k \ge 2$, $u \in V$ is {\em $(v_1,\ldots,v_k)$-forced},if $u \in D$ for every e.d.s.\ $D$ of $G$ with $v_1,\ldots,v_k \in D$, and correspondingly, $u \in V$ is {\em $(v_1,\ldots,v_k)$-excluded} if $u \notin D$ for such e.d.s.\ $D$, and $G$ can be reduced by the same principle.

\medskip

In this manuscript, we solve some cases in polynomial time by the following approach: Assume that $G$ has an e.d.s.\ $D$. Then
for any vertex $v \in V$, $|N[v] \cap D|=1$. Assume that $v \in D$ for a vertex $v \in V$, and let $N_i$, $0 \le i \le k$, denote the distance levels of $v$ in $G$ (in particular, $N_0=\{v\}$). Since $G$ is bipartite, every $N_i$ is independent. Since we assume that $v \in D$ and $D$ is an e.d.s.\ of $G$ (i.e., the closed neighborhoods of the $D$-vertices form a partition of $V$), all vertices in $N_1 \cup N_2$ are $v$-excluded, and thus we have:
\begin{equation}\label{N1N2Dempty}
D \cap (N_1 \cup N_2) = \emptyset.
\end{equation}

Here are examples when $v \in D$ is impossible as well as examples for $v$-forced vertices:
\begin{Claim}\label{vforcedvnotinD}
Assume that $v \in D$. Then:
\begin{itemize}
\item[$(i)$] If for some $x \in N_2$, $N(x) \cap N_3 = \emptyset$ then $G$ has no e.d.s.\ $D$ with $v \in D$.

\item[$(ii)$] If for some $x \in N_2$, $N(x) \cap N_3 = \{y\}$ then $y$ is $v$-forced.

\item[$(iii)$] If for $y \in N_3$, $N(y) \cap N_4 = \emptyset$ then $y$ is $v$-forced.

\item[$(iv)$] If for $y_1,y_2 \in N_3$, $y_1 \neq y_2$, $N(y_1) \cap N_4 = N(y_2) \cap N_4 = \emptyset$ and $N(y_1) \cap N(y_2) \cap N_2 \neq \emptyset$ then $G$ has no e.d.s. $D$ with $v \in D$.
\end{itemize}
\end{Claim}

Thus, after reducing $G$ by such $v$-forced vertices, for finding an e.d.s.\ $D$ with $v \in D$, we can assume:
\begin{equation}\label{xN3N4neighb}
\mbox{For every } y \in N_3, N(y) \cap N_4 \neq \emptyset.
\end{equation}

More generally, for $i \geq 1$, let us write

\begin{itemize}
\item[ ] $N_i^* = \{x \in N_i: N(x) \cap N_{i+1} \neq \emptyset\}$, and
\item[ ] $N_i^0 = \{x \in N_i: N(x) \cap N_{i+1} = \emptyset\}$,
\end{itemize}
i.e., $\{N_i^*,N_i^0\}$ is a partition of $N_i$.

\medskip

For $y_i \in D \cap N_3$, let $x_i \in N_2$ and $z_i \in N_4$ with $x_iy_i \in E$ and $y_iz_i \in E$ for $i \in \{1,2\}$ (recall (\ref{xN3N4neighb})). Clearly, for every pair $y_1,y_2 \in D \cap N_3$, $y_1 \neq y_2$, we have:
\begin{equation}\label{2P3N2N3N4}
x_1,x_2,y_1,y_2,z_1,z_2 \mbox{ induce a } 2P_3.
\end{equation}

\begin{Claim}\label{S224threeDinN3}
Assume that $D \cap N_3=\{y_1,\ldots,y_k\}$ for $k \ge 3$ $($recall $(\ref{xN3N4neighb}))$. Then:
\begin{itemize}
\item[$(i)$] If at least three vertices in $N(D \cap N_3) \cap N_2$, say $x_1,x_2,x_3$, have private neighbors in $N_1$ then there is an $S_{4,4,4}$ in $G$.
\item[$(ii)$] If all of $x_1,x_2,x_3$ have a common neighbor in $N_1$ then there is an $S_{3,3,3}$ in $G$.
\item[$(iii)$] If neither $(i)$ nor $(ii)$ appears then there is an $S_{2,4,4}$ or $S_{3,3,5}$ in $G$.
\end{itemize}
\end{Claim}

\noindent
{\em Proof.}
$(i)$: Let $u_i \in N_1$ be the private neighbors of $x_i$, $i \in \{1,2,3\}$. Then $v,u_1,x_1,y_1,z_1$, $u_2,x_2,y_2,z_2,u_3,x_3,y_3,z_3$ (with center $v$) induce an $S_{4,4,4}$.

\noindent
$(ii)$: Let $u \in N_1$ be a common neighbor of $x_i$, $i \in \{1,2,3\}$. Then $u,x_1,y_1,z_1$, $x_2,y_2,z_2,x_3,y_3,z_3$ (with center $u$) induce an $S_{3,3,3}$.

\noindent
$(iii)$: Assume that not all of $x_1,x_2,x_3$ have private neighbors in $N_1$ and there is no common neighbor of $x_1,x_2,x_3$ in $N_1$ (since neither $(i)$ nor $(ii)$ appears). Without loss of generality, let
$u \in N_1$ be a common neighbor of $x_1,x_2$ such that $ux_3 \notin E$, and let $u' \in N_1$ with $u'x_3 \in E$.

If $u'x_1 \in E$ and $u'x_2 \notin E$ then $x_1,y_1,z_1,u,x_2,y_2,z_2,u',x_3,y_3,z_3$ (with center $x_1$) induce an $S_{2,4,4}$, and similarly,
if $u'x_2 \in E$ and $u'x_1 \notin E$ then $x_2,y_2,z_2,u,x_1,y_1,z_1,u',x_3,y_3,z_3$ (with center $x_2$) induce an $S_{2,4,4}$.

Finally, if $u'x_1 \notin E$ and $u'x_2 \notin E$ then $u,x_1,y_1,z_1,x_2,y_2,z_2,v,u',x_3,y_3,z_3$ (with center $u$) induce an $S_{3,3,5}$. Thus, Claim \ref{S224threeDinN3} is shown.
$\diamond$

\medskip

Recall that we assume that $v \in D$ for a vertex $v \in V$, and $N_i$, $1 \le i \le k$ are the distance levels of $v$.
Let $$G_i:=G[\{v\} \cup N_1 \cup \ldots \cup N_i]$$ and assume that $$D_i:=D \cap (\{v\} \cup N_1 \cup \ldots \cup N_i)$$ is known as an e.d.s.\ for $G_{i-1}$.
In particular, $D_0=D_1=D_2=\{v\}$. Note that for $i \ge 3$, $D_i$ additionally dominates parts of $N_i \cup N_{i+1}$.

Let $$N'_i:=N_i \setminus (D_i \cup N(D_i)).$$

Recall that for any $d,d' \in D$, $d \neq d'$, $d_G(d,d') \ge 3$. Thus, for possible $D$-candidates from $N_{i+1}$, we have to exclude $N(D_i) \cup N^2(D_i)$ from $N_{i+1}$ (recall the notion of $N^2(D_i)$ as in the Introduction), i.e., $$W_{i+1}:=N_{i+1} \setminus (N(D_i) \cup N^2(D_i)).$$
The collection of possible $D$-candidates in $W_{i+1}$ has to dominate $N'_i$.

Thus, for constructing $D_{i+1}$ from $D_i$, for a possible subset $Z_{i+1} \subset W_{i+1}$ such that $Z_{i+1}$ dominates $N'_i$ (following the e.d.s.\ property), we have $D_{i+1}=D_i \cup Z_{i+1}$.

Finally, if $N_k$ is the last distance level of $v$ then $N_k \setminus N(D_{k-1}) \subset D$ (if there exists an e.d.s.\ $D$ with $v \in D$).

\medskip

The stepwise construction of $D_{i+1}$ from $D_i$ is possible e.g.\ when candidates are forced (and the reduction follows): Recall that $D_1=\{v\}$ is an e.d.s.\ for $G_1$. Similarly as for $v$-forced vertices, a vertex $u \in W_{i+1}$ is {\em $D_i$-forced} if $u \in D$ for every e.d.s.\ $D$ of $G$ with $D_i \subset D$, $D_i$ being an e.d.s.\ for $G_{i-1}$.

As in Claim \ref{vforcedvnotinD} $(iii)$ and (\ref{xN3N4neighb}), a non-excluded vertex $x \in N_{i+1}$ (with respect to $D_i$) with neighbor in $N'_i$ is $D_i$-forced if $N(x) \cap N_{i+2} = \emptyset$. Thus we can assume that $N(x) \cap N_{i+2} \neq \emptyset$, i.e., $x \in N_{i+1}^*$.

\medskip

One of the helpful arguments is that in some cases, $v$ has only a fixed number of distance levels; for instance, this is the case for $P_k$-free bipartite graphs (e.g., for $P_5$-free bipartite graphs, we have $N_4=\emptyset$) as well as, more generally, for $\ell P_k$-free bipartite graphs (but the complexity of ED is still open for $P_8$-free bipartite graphs and for $\ell P_5$-free bipartite graphs).

A more general case is when for all $i \ge k$, every $x \in N_i$ has at most one neighbor in $N_{i+1}$; for instance, this is the case for $S_{1,1,k}$-free bipartite graphs (for which the complexity of ED is still open).

Another more general case is when an e.d.s.\ $D$ in $G$ has only polynomially many subsets in $N_i$, $i \ge 3$.
If $v$ has a fixed number of $k$ distance levels $N_i$, then, starting with $D_1=\{v\}$, for $i:=1$ to $k$, we can produce a polynomial number of $D_i$. This can be done for $P_7$-free bipartite graphs and for $\ell P_4$-free bipartite graphs (see Section \ref{P7free}).

\medskip

If the number of distance levels $N_i$ of $v$ is not fixed, it leads to a polynomial time algorithm for ED when e.g.\ $D$ has polynomially many subsets in $N_3$ and
for each $D_i$, $i \ge 2$, the candidates for $D_{i+1}$ are $D_i$-forced. This can be done e.g.\ for $S_{2,2,4}$-free bipartite graphs (see Section \ref{S224free}).
A more special case is when for all distance levels $N_i$, $i \ge k$ for a fixed number $k$, there is at most one neighbor of $x \in N_i$ in $N_{i+1}$, and the number of e.d.s.\ candidates for $G_{k-1}$ is polynomial. This can be done for $S_{1,2,4}$-free bipartite graphs (see Section \ref{S224free}).

\subsection{ED in polynomial time for $H$-free bipartite graphs when $H=P_7$ or $H= \ell P_4$}\label{P7free}

Recall that ED is \NP-complete for $P_7$-free graphs but polynomial for $P_6$-free graphs (see e.g.\ \cite{BraMos2016}). Moreover, for $P_5$-free bipartite graphs $G=(V,E)$, for every $v \in V$, $N_4=\emptyset$ and thus, $D=\{v\} \cup N_3$ is unique (if it is really an e.d.s.) and thus, ED can be solved in linear time for $P_5$-free bipartite graphs; actually, ED is done in linear time for $P_5$-free graphs \cite{BraMos2016}.
The subsequent lemma implies further polynomial cases for ED:

\begin{lemma}[\cite{BraGia2014,BraGiaMil2018}]\label{P2cojoin}
If ED is solvable in polynomial time for $F$-free graphs then ED is solvable in polynomial time for $(P_2+F)$-free graphs.
\end{lemma}

This clearly implies the corresponding fact for $(P_1+F)$-free graphs.
By Theorem \ref{cwdHfrbipgr}, the clique-width of $P_7$-free bipartite graphs is unbounded.

Recall that for graph $G=(V,E)$, the distance $d_G(a,b)$ between two vertices $a,b$ of $G$ is the number of edges in a shortest path between $a$ and $b$ in $G$.

\begin{theorem}[\cite{BacTuz1990/2}]\label{theoB-T}
Every connected $P_t$-free graph $G=(V,E)$ admits a vertex $v \in V$ such that $d_G(v,u) \leq \bigl\lfloor t/2 \bigr\rfloor$ for every $u \in V$.
\end{theorem}

\begin{theorem}\label{EDP7frbipgr}
For $P_7$-free bipartite graphs, ED is solvable in time $O(n^4)$.
\end{theorem}

\noindent
{\bf Proof.}
Let $G=(V,E)$ be a connected $P_7$-free bipartite graph. Recall that we assume that for an e.d.s.\ $D$ of $G$, $v \in D$ and $N_i$, $i \ge 1$, denote the distance levels of $v$ in $G$; since $G$ is bipartite, $N_i$ is independent for every $i \ge 1$.
By Theorem~\ref{theoB-T}, there is a vertex $v_0$ whose distance levels $N_k(v_0)$, $k \ge 4$, are empty. Moreover, for an e.d.s.\ $D$ of $G$, either $v_0 \in D$ or there is a neighbor $v$ of $v_0$ such that $v \in D$. Thus, since $G$ is $P_7$-free, for every $v \in N[v_0]$, $N_5=\emptyset$.

Thus, for every $v \in N[v_0]$, we can check whether $v$ is part of an e.d.s.\ $D$ of $G$.
By (\ref{N1N2Dempty}), for $v \in D$, we have $D \cap (N_1 \cup N_2) = \emptyset$.
By Claim \ref{vforcedvnotinD} $(iii)$, if for $y \in N_3$, $N(y) \cap N_4 =\emptyset$ then $y$ is $v$-forced, and in particular, if $N_4 =\emptyset$ then we have to check whether $\{v\} \cup N_3$ is an e.d.s. (for example, this has to be done for $v=v_0$).

Thus, we can reduce the graph (recall Claims \ref{forcedreduction} and \ref{vforcedreduction}) and from now on,
by (\ref{xN3N4neighb}) assume that for every $y \in N_3$, $N(y) \cap N_4 \neq \emptyset$. If there are two such vertices $y_1,y_2 \in N_3 \cap D$ (with neighbors  $x_i \in N_2$ and $z_i \in N_4$, $i \in \{1,2\}$) then, since $x_1,y_1,z_1,x_2,y_2,z_2$ induce a $2P_3$ in $G$, $G$ would contain a $P_7$. Thus, after the reduction, $|D \cap N_3| = 1$; let $d_1 \in D \cap N_3$. Then again, since $N_5=\emptyset$, every $z \in N_4 \setminus N(d_1)$ is $(v,d_1)$-forced.

The algorithm has the following steps (according to Section \ref{generalapproach}):
\begin{itemize}
\item[(A)] Find a vertex $v_0$ in $G$ with $N_4(v_0)=\emptyset$.
\item[(B)] For each $v \in N[v_0]$, check whether there is an e.d.s.\ $D$ of $G$ with $v \in D$, as follows:
\begin{itemize}
\item[(B.$1$)] Add all vertices $y \in N_3$ with $N(y) \cap N_4=\emptyset$ to the initial $D=\{v\}$ and reduce $G$ correspondingly. If $D$ has a contradiction to the e.d.s.\ property then $G$ has no e.d.s.\ $D$ with $v \in D$.
\item[(B.$2$)] For the reduced graph $G'$, check for any vertex $y \in N_3$ with $N(y) \cap N_4 \neq \emptyset$ whether $D \cup \{y\} \cup (N_4 \setminus N(y))$ is an e.d.s.\ of $G'$.
\end{itemize}
\end{itemize}

Clearly, this can be done in time $O(n^4)$. Thus, Theorem \ref{EDP7frbipgr} is shown.
\qed

\medskip

For $\ell P_4$-free bipartite graphs, $\ell \ge 2$, we again show that ED is solvable in polynomial time (recall that for $\ell=1$, it is done already).

\begin{theorem}[\cite{Aleks1991,BalYu1989,Farbe1989,FarHujTuz1993,PauUng1959,Prisn1995}]\label{theoFarber}
$2K_2$-free graphs with $n$ vertices have at most $n^2$ maximal independent sets, and these can be computed in time $O(n^4)$. More generally, for fixed $\ell \ge 2$,
$\ell K_2$-free graphs with $n$ vertices have at most $n^{2 \ell -2}$ maximal independent sets, and these can be computed in polynomial time $O(n^{2 \ell})$.
\end{theorem}

\begin{theorem}\label{theoWEDlP4frbip}
For $\ell P_4$-free bipartite graphs, for any fixed $\ell$, ED is solvable in polynomial time for every fixed $\ell \ge 2$.
\end{theorem}

\noindent
{\bf Proof.} The proof is similar to the approach in Section \ref{generalapproach}; recall that the ED problem can be reduced to prime graphs. In particular,
since $G$ is $\ell P_4$-free, for every possible $v \in D$ we have $N_k=\emptyset$ for every $k \geq 5 \ell - 2$. Thus, the number of distance levels is finite.

Moreover, let $H:=G^2[N_3]$. We first claim that $H$ is $\ell K_2$-free: Suppose to the contrary that there is an $\ell K_2$ in $H$, say $a_1b_1,a_2b_2,\ldots,a_{\ell}b_{\ell} \in E(H)$, $a_i, b_i \in N_3$. Thus, since $N_3$ is independent, $d_G(a_i,b_i)=2$, i.e., $a_i$ and $b_i$ have a common neighbor $c_i \in N_2 \cup N_4$, $i \in \{1,\ldots,\ell\}$. Since $a_1b_1,a_2b_2,\ldots,a_{\ell}b_{\ell}$ should be an $\ell K_2$ in $H$, the pairwise $G$-distance between $a_ib_i$ and $a_jb_j$ is at least 3 for every $i \neq j$. Thus, there is no common $G$-neighbor between $a_ib_i$ and $a_jb_j$, and there are no $G$-edges between them.

Since $G$ is prime (and $\{a_i,b_i\}$ is no module), there is a vertex $d_i \in N_2 \cup N_4$ distinguishing $a_i$ and $b_i$, say $d_ia_i \in E(G)$ and $d_ib_i \notin E(G)$, and now, $a_i,b_i,c_i,d_i$ induce a $P_4$ in $G$. By the distance properties and the assumption that $a_1b_1,a_2b_2,\ldots,a_{\ell}b_{\ell}$ should be an $\ell K_2$ in $H$, we have
$\{a_i,b_i,c_i,d_i\} \cap \{a_j,b_j,c_j,d_j\} = \emptyset$ and there are no $G$-edges between them, but now, $a_1,b_1,c_1,d_1,\ldots,a_{\ell},b_{\ell},c_{\ell},d_{\ell}$ induce an $\ell P_4$ in $G$, which is a contradiction. Thus, $H$ is $\ell K_2$-free.

Since $D \cap N_3$ is a maximal independent set in $H$, it follows by Theorem~\ref{theoFarber} that there are polynomially many such possible subsets $D \cap N_3$.

Then, starting with one of the possible subsets $D \cap N_3$, it can be continued for the fixed number of remaining distance levels as in the approach in Section \ref{generalapproach}. Thus, Theorem \ref{theoWEDlP4frbip} is shown.
\qed

\begin{corollary}\label{coroWEDlP4frbip}
For every fixed $\ell$, each $\ell P_4$-free bipartite graph has a polynomial number of possible e.d.s., and it can be computed in polynomial time.
\end{corollary}

\subsection{ED for $S_{2,2,4}$-free bipartite graphs in polynomial time}\label{S224free}

In this section, we generalize the ED approach for $P_7$-free bipartite graphs. Recall that the clique-width of $S_{1,2,3}$-free bipartite graphs is bounded and the clique-width of $S_{1,2,4}$-free bipartite graphs as well as of $S_{2,2,3}$-free bipartite graphs is unbounded.

As usual, we check for every $v \in V$ whether $v$ is part of an e.d.s.\ $D$ of $G$.
Let $N_i$, $i \geq 1$, denote the distance levels of $v$ in $G$; since $G$ is bipartite, every $N_i$ is an independent vertex subset.
Recall by (\ref{N1N2Dempty}) that $D \cap (N_1 \cup N_2) = \emptyset$, and by (\ref{xN3N4neighb}), for every $y \in N_3$, $N(y) \cap N_4 \neq \emptyset$, i.e., subsequently we consider only $D$-candidates in $N_3$ which are not $v$-forced.

\medskip

The following is a general approach which will be used for $S_{2,2,k}$-free bipartite graphs, $k \in \{2,3,4\}$,
and for $S_{1,2,4}$-free bipartite graphs:

Recall that $G_i=G[\{v\} \cup N_1 \cup \ldots \cup N_i]$ and assume that $D_i:=D \cap (\{v\} \cup N_1 \cup \ldots \cup N_i)$, is an e.d.s.\ for $G_{i-1}$.
Moreover, recall $N'_i=N_i \setminus (D_i \cup N(D_i))$ and $W_{i+1}=N_{i+1} \setminus (N(D_i) \cup N^2(D_i))$.
Clearly, if for $x \in N'_i$, there is no neighbor of $x$ in $W_{i+1}$ then there is no such e.d.s.\ $D$, and if
$|N(x) \cap W_{i+1}|=1$ then the corresponding neighbor of $x$ in $W_{i+1}$ is $D_i$-forced. Now assume that $|N(x) \cap W_{i+1}| \ge 2$ for $i \ge k$.

\begin{Claim}\label{S22ktwoN(x)inNi+1}
If for $k \ge 2$, $G$ is $S_{2,2,k}$-free bipartite and for $x \in N'_i$, $i \ge k$, $|N(x) \cap W_{i+1}| \ge 2$ then for the $D$-vertex $y \in N(x) \cap W_{i+1}$ which dominates $x$, we have $N(y) \cap N_{i+2} \subset N(y') \cap N_{i+2}$ for every $y' \in N(x) \cap W_{i+1}$, $y' \neq y$.
\end{Claim}

\noindent
{\em Proof.} Let $y \in N(x) \cap W_{i+1}$ be the $D$-vertex which dominates $x$, and let $y' \in N(x) \cap W_{i+1}$, $y' \neq y$.
Let $z \in N(y) \cap N_{i+2}$ be any neighbor of $y$ in $N_{i+2}$. Since $y \in D$, $y'$ has to be dominated by a neighbor $z' \in N_{i+2} \cap D$, and since for a shortest path $(x,x_{i-1},x_{i-2},\ldots,v)$, $x_j \in N_j$, between $x$ and $v$, the subgraph induced by vertices $x,y,z,y',z',x_{i-1},x_{i-2},\ldots,v$ (with center $x$) do not contain an induced $S_{2,2,k}$, we have $y'z \in E$. Thus, $N(y) \cap N_{i+2} \subset N(y') \cap N_{i+2}$, and Claim \ref{S22ktwoN(x)inNi+1} is shown.
$\diamond$

\begin{theorem}\label{theoS224}
For $S_{2,2,4}$-free bipartite graphs, ED is solvable in time $O(n^6)$.
\end{theorem}

\noindent
{\bf Proof.} Let $G=(V,E)$ be an $S_{2,2,4}$-free bipartite graph.
Recall that, as in Section \ref{generalapproach}, for $i \ge 3$, we denote $G_{i-1}=G[\{v\} \cup N_1 \cup \ldots \cup N_{i-1}]$, $D_i=D \cap (\{v\} \cup N_1 \cup \ldots \cup N_i)$, $N'_i=N_i \setminus (D_i \cup N(D_i))$, and $W_{i+1}=N_{i+1} \setminus (N(D_i) \cup N^2(D_i))$.

By the e.d.s.\ property, the collection of possible $D$-candidates from $W_{i+1}$ has to dominate $N'_i$.
Thus, for constructing $D_{i+1}$ from $D_i$, for a possible subset $Z_{i+1} \subseteq W_{i+1}$ such that $Z_{i+1}$ dominates $N'_i$, we have $D_{i+1}=D_i \cup Z_{i+1}$.

First let us see how many subsets $Q$ of $N_3$ are candidates for $D \cap N_3 = Q$, i.e., for $D_3 = Q \cup \{v\}$.

\begin{Claim}\label{S224Claim2}
For any $y \in D \cap N_3$, the remaining $D$-vertices in $D \cap N_3 \setminus \{y\}$ are $(v,y)$-forced.
\end{Claim}

\noindent
{\em Proof.} Let us fix any possible vertex $y \in D \cap N_3$, and let $Y := N[y]$, i.e., $Y := \{y\} \cup (N(y) \cap N_2) \cup (N(y) \cap N_4)$, since $N_2,N_3$, and $N_4$ are independent, let $M_2 := N_2 \setminus N(y)$, $M_3 := N_3 \setminus N[Y]$, and $M_4 := N_4 \setminus N(y)$; let $x \in N_2 \cap N(y)$ and $z \in N_4 \cap N(y)$, and let $u \in N_1 \cap N(x)$.
By construction and by the e.d.s.\ property, the remaining $D$-vertices in $D \cap N_3 \setminus \{y\}$ are in $M_3$ and their neighbors are in $M_2 \cup M_4$.

Then let us fix any $x' \in M_2$. Since we assumed that $y \in D$, vertex $x'$ has to be dominated by some vertex in $M_3$.

Clearly, if there is no neighbor of $x'$ in $M_3$ then there is no such e.d.s.\ $D$, and if $|N(x') \cap M_3|=1$ then the corresponding neighbor of $x'$ in $M_3$ is $(v,y)$-forced. Now assume that $|N(x') \cap M_3| \geq 2$.

Let us show that, similarly as for Claim \ref{S22ktwoN(x)inNi+1}, for the $D$-vertex $y^* \in M_3 \cap N(x')$ which dominates $x'$, and for every $y' \in M_3 \cap N(x')$, $y' \neq y^*$, we have:
\begin{equation}\label{S224frN4neighbinclusion}
N(y^*) \cap M_4 \subset N(y') \cap M_4.
\end{equation}

\noindent
{\em Proof of $(\ref{S224frN4neighbinclusion})$.} Let $y^* \in M_3 \cap N(x)$ be the $D$-vertex which dominates $x'$, and let $y' \in M_3 \cap N(x')$, $y' \neq y^*$. Let $z^* \in N(y^*) \cap M_4$ be any neighbor of $y^*$ in $M_3$. Since $y^* \in D$, $y'$ has to be dominated by a neighbor $z' \in M_4 \cap D$.

First assume that $x'u \in E$. Then, since $x',y^*,z^*,y',z',u,x,y,z$ (with center $x'$) does not induce an $S_{2,2,4}$, we have $y'z^* \in E$. Thus $N(y^*) \cap M_4 \subset N(y') \cap M_4$.

Now assume that $x'u \notin E$; let $u' \in N_1$ with $u'x' \in E$. Without loss of generality, let $u'x \notin E$ (else $u$ can be replaced by $u'$ in the previous argument). Then again, since $x',y^*,z^*,y',z',u',v,u,x$ (with center $x'$) does not induce an $S_{2,2,4}$, we have $y'z^* \in E$. Thus $N(y^*) \cap M_4 \subset N(y') \cap M_4$, i.e., the assertion (\ref{S224frN4neighbinclusion}) is shown.
$\diamond$

\medskip

Then (\ref{S224frN4neighbinclusion}) implies that the candidates for $D \cap N_3 \setminus \{y\}$ are $(v,y)$-forced: For any $x' \in M_2$ with $|N(x') \cap M_3| \geq 2$, the vertex $y^* \in M_3 \cap N(x')$ with $N(y^*) \cap M_4 \subset N(y') \cap M_4$ for every $y' \in M_3 \cap N(x')$, $y' \neq y$, is $(v,y)$-forced.

Summarizing: let $y \in D \cap N_3$; then every $x' \in M_2$ has to be dominated by a vertex of $N_3$, say $y(x')$, and there is at most one candidate for such a vertex (if there is no candidate, then there is no e.d.s. $D$ with $v,y \in D$); in particular $y(x')$ can be computed in polynomial time; then $y \in D$ if and only if $Q_y = D \cap N_3$ where
$Q_y = \{y\} \cup \{y(x'): x' \in M_2\}$. Thus, Claim \ref{S224Claim2} is shown.
$\diamond$

\begin{Claim}\label{S224Claim3}
There are at most $n$ subsets $Q$ of $N_3$ which are candidates for $D \cap N_3 = Q$, i.e., for $D_3 = Q \cup \{v\}$.
\end{Claim}

\noindent
{\em Proof.} 
According to the last lines of the proof of Claim \ref{S224Claim2}, for each vertex $y \in N_3$ one can compute in polynomial time a set $Q_y \subseteq N_3$ such that $y \in D$ if and only if $D \cap N_3 = Q_y$ (or one can show that there is no e.d.s. $D$ with $v,y \in D$), so that there are at most $|N_3|$ subsets $Q$ of $N_3$ which are candidates for $D \cap N_3 = Q$.
$\diamond$

\medskip

Then let us consider the set $D_4$.

\begin{Claim}\label{S224Claim4}
There are at most $n^2$ subsets $Q'$ of $N_3 \cup N_4$ which are candidates for $D \cap (N_3 \cup N_4) = Q'$, i.e., for $D_4 = Q' \cup \{v\}$.
\end{Claim}

\noindent
{\em Proof.} By Claim \ref{S224Claim3} there are at most $n$ subsets $Q$ of $N_3$ which may be candidates for $D \cap N_3 = Q$, i.e., for $D_3 = Q \cup \{v\}$. Then for each such candidate for $D_3$, one can iterate the approach in the proof of Claim \ref{S224Claim2} which leads to Claim \ref{S224Claim3}, in order to construct candidates for $D_4$ from $D_3$ [according to the general approach, i.e., for a possible subset $Z_4 \subset N_4 \setminus N(D_3)$ such that $Z_4$ dominates $N_3 \setminus (D_3 \cup N(D_3))$ (following the e.d.s.\ properties), we have $D_4=D_3 \cup Z_4$].
$\diamond$

\medskip

Then let us consider the sets $D_i$ for $i \geq 4$. Recall that $N'_i=N_i \setminus (D_i \cup N(D_i))$ and $W_{i+1} = N_{i+1} \setminus (N(D_i) \cup N^2(D_i))$. By Claim \ref{S22ktwoN(x)inNi+1}, we have:
\begin{Claim}\label{S224twoN(x)inNi+1}
If for $x \in N'_i$, $i \ge 4$, $|N(x) \cap W_{i+1}| \ge 2$ then for the $D$-vertex $y \in W_{i+1} \cap N(x)$ which dominates $x$, we have $N(y) \cap N_{i+2} \subset N(y') \cap N_{i+2}$ for every $y' \in N(x) \cap W_{i+1}$, $y' \neq y$.
\end{Claim}

By Claim \ref{S224Claim4}, we can start by checking at most $n^2$ subsets $Q'$ of $N_3 \cup N_4$ which are candidates for $D \cap (N_3 \cup N_4) = Q'$, i.e., for $D_4 = Q' \cup \{v\}$. Then, according to Claim \ref{S224twoN(x)inNi+1}, for each such candidate for $D_4$, there is just one possible extension for $D_i$ with $i \geq 5$.
Then by Claim \ref{S224twoN(x)inNi+1}, this leads to another forced condition: For any $x \in N'_i$, the vertex $y \in N(x) \cap W_{i+1}$ with $N(y) \cap N_{i+2} \subset N(y') \cap N_{i+2}$ for every $y' \in N(x) \cap W_{i+1}$, $y' \neq y$, is $D_i$-forced.

Checking the neighborhood inclusion in Claim \ref{S224twoN(x)inNi+1} can be done in time $O(n^2)$ for each vertex $y \in W_{i+1} \cap N(x)$. Finally, since altogether, there are at most $n^3$ possible e.d.s.\ in $G$ (by adding the starting vertex $v$), ED is solvable in time $O(n^6)$ for $S_{2,2,4}$-free bipartite graphs, and Theorem \ref{theoS224} is shown.
\qed

\medskip

Subsequently, we improve the time bound for some subclasses of $S_{2,2,4}$-free bipartite graphs.

\begin{corollary}\label{theoS223}
For $S_{2,2,3}$-free bipartite graphs, ED is solvable in time $O(n^5)$.
\end{corollary}

\noindent
{\bf Proof.} For $S_{2,2,3}$-free bipartite graphs, Claim \ref{S224twoN(x)inNi+1} is already available for $x \in N'_i$, $i \ge 3$.
\qed

\medskip

For the special case of $S_{2,2,2}$-free bipartite graphs, Claim \ref{S224twoN(x)inNi+1} is already available for $x \in N'_i$, $i \ge 2$.
Thus, the assumption that $v \in D$ and the distance levels of $v$ imply that every other vertex in $D$ is $v$-forced. Then each vertex of $G$ is contained in at most one e.d.s.\ of $G$.

\begin{corollary}
Every connected $S_{2,2,2}$-free bipartite graph contains at most $n$ e.d.s.\ and these e.d.s.\ can be computed in time $O(n^3)$.
\end{corollary}

For $S_{1,2,4}$-free bipartite graphs, Theorem \ref{theoS224} is available. For the algorithmic approach, we can use a more special version:
Without loss of generality, let us assume that $G$ is prime (recall the corresponding comment in the Introduction).

\begin{Claim}\label{Nkatmost1Nk+1}
Let $G$ be a prime $S_{1,2,4}$-free bipartite graph with e.d.s.\ $D$ and let $v \in D$. Then, for $k \geq 5$, each vertex of $N_k$ has at most one neighbor in $N_{k+1}$.
\end{Claim}

\noindent
{\em Proof.}
Without loss of generality, let $u_5 \in N_5$, and let $v,u_1,u_2,u_3,u_4,u_5$ be a shortest path from $v$ to $u_5$.
Suppose to the contrary that $u_5$ has two neighbors, say $q_1,q_2$ in $N_6$. Then, since $G$ is prime, there exists a vertex $y \in N_5 \cup N_7$ distinguishing $q_1,q_2$, say $yq_1 \in E$ and $yq_2 \notin E$.

If $y \in N_7$ or $y \in N_5$ and $yu_4 \notin E$ then $u_5,q_2,q_1,y,u_4,u_3,u_2,u_1$ (with center $u_5$) induce an $S_{1,2,4}$, which is a contradiction.
Thus, $y \in N_5$ and $yu_4 \in E$ but then $u_4,y,u_5,q_2,u_3,u_2,u_1,v$ (with center $u_4$) induce an $S_{1,2,4}$, which is again a contradiction.

Analogously, for $u_k \in N_k$, $k \geq 6$, with two neighbors $q_1,q_2$ in $N_{k+1}$, $G$ contains an $S_{1,2,4}$.
Thus, Claim \ref{Nkatmost1Nk+1} is shown.
$\diamond$

\begin{Claim}\label{S124ynotjoinN2N3}
Let $G$ be a prime $S_{1,2,4}$-free bipartite graph with e.d.s.\ $D$ and let $v \in D$. If for $y \in D \cap N_3$, $N(y) \cap N_2 \subset N_2$ then for every $x' \in N_2 \setminus N(y)$, $x'$ has at most one neighbor in $N_3 \setminus N(N(y))$.
\end{Claim}

\noindent
{\em Proof.}
Let $x \in N_2 \cap N(y)$, $z \in N_4 \cap N(y)$ and $u \in N(x) \cap N_1$. Since $N(y) \cap N_2 \subset N_2$, there is a vertex $x' \in N_2 \setminus N(y)$.
Since $x'$ has to be dominated by a $D$-vertex in $N_3$, $x'$ must have a neighbor in $N_3 \setminus N(N(y))$ by the e.d.s.\ property.

Suppose to the contrary that $|N(x') \cap (N_3 \setminus N(N(y)))| \ge 2$. Then there is a $D$-vertex
$y^* \in N(x') \cap (N_3 \setminus N(N(y)))$, and let $y' \in N(x') \cap (N_3 \setminus N(N(y)))$ be a second neighbor of $x'$. Now, $y'$ has to be dominated by a $D$-vertex $z' \in D \cap N_4$. Clearly, $\{x,z\} \cojoin \{y',y^*\}$.

Since $x',y^*,y',z',u,x,y,z$ (with center $x'$) do not induce an $S_{1,2,4}$, we have $x'u \notin E$ and in general, $x$ and $x'$ do not have a common neighbor in $N_1$. Thus, let $u' \in N_1$ be a neighbor of $x'$ but now $x',y^*,y',z',u',v,u,x$ (with center $x'$) induce an $S_{1,2,4}$, which is a contradiction.
Thus, Claim \ref{S124ynotjoinN2N3} is shown.
$\diamond$

\begin{corollary}\label{WEDS124frbip}
For every connected $S_{1,2,4}$-free bipartite graph, ED is solvable in time $O(n^4)$.
\end{corollary}

\section{ED for $H$-free chordal bipartite graphs with degree $\le$ 3}

Recall that a bipartite graph $G$ is chordal bipartite if $G$ is $C_{2k}$-free bipartite for every $k \ge 3$.
In \cite{DabLozZam2012}, it is shown that the clique-width of $A$-free chordal bipartite graphs is at most 6.
In \cite{BerBraEng2015}, it is shown that a graph is chordal bipartite and its mirror is chordal bipartite if and only if it is $(3P_2,C_6,C_8)$-free bipartite.
These graphs were called {\em auto-chordal bipartite} graphs in \cite{BerBraEng2015}. Thus, ED is solvable in polynomial time for auto-chordal bipartite graphs.

\medskip

Recall that ED for $G$ can be solved by MWIS for $G^2$. The subgraph $A_4$ has five vertices, say $v_1,\ldots,v_5$ such that $v_1,\ldots,v_4$ induce a $C_4$ and $v_5$ is adjacent to exactly one of $v_1,\ldots,v_4$, say $v_5v_3 \in E$. In \cite{BraLozMos2010}, it is shown that MWIS for (hole,$A_4$)-free graphs is solvable in polynomial time.

\begin{figure}[ht]
  \begin{center}
   \epsfig{file=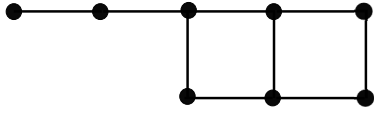}
   \caption{extended domino $H_4$}
   \label{extdomH4}
  \end{center}
\end{figure}

If we restrict chordal bipartite graphs to degree at most 3 then we obtain the following result:

\begin{theorem}\label{EDchordbipgrdeg3}
Let $G=(X,Y,E)$ be a chordal bipartite graph with vertex degree at most $3$. Then:
\begin{itemize}
\item[$(i)$] $G^2$ is hole-free.
\item[$(ii)$] If $G$ is $H_4$-free then $G^2$ is $A_4$-free.
\end{itemize}
\end{theorem}

\noindent
{\bf Proof.} $(i)$: Suppose to the contrary that there is a hole $(v_1,\ldots,v_k)$, $k \ge 5$, in $G^2$. Without loss of generality, let $v_1 \in X$.
If $d_G(v_i,v_{i+1})=2$ (modulo $k$) for all $i \in \{1,\ldots,k\}$  then all common neighbors of $v_i,v_{i+1}$ are in $Y$ and thus, there is a hole in $G$, which is a contradiction since $G$ is chordal bipartite.
Thus, at least one of the pairs $v_i,v_{i+1}$ has distance 1; without loss of generality let $v_1v_2 \in E$. Then $d_G(v_k,v_1)=2$ and $d_G(v_2,v_3)=2$; let $x_2 \in X$ be a common neighbor of $v_2,v_3$ and let $y_k \in Y$ be a common neighbor of $v_k,v_1$. If $x_2y_k \notin E$ then it leads to a hole in $G$ which is impossible. Thus, $x_2y_k \in E$ but now, the degrees of $x_2$ and $y_k$ are 3, and thus, by the degree bound, $x_2$ and $y_k$ have no other neighbors. Now the cycle $(x_2,v_3,\ldots,v_k,y_k)$ leads to a hole in $G$, which is a contradiction. Thus, $G^2$ is hole-free.

\medskip

\noindent
$(ii)$: Suppose to the contrary that $v_1,\ldots,v_5$ induce an $A_4$ in $G^2$ with $C_4$ $(v_1,v_2,v_3,v_4)$ in $G^2$ and vertex $v_5$ with $v_5v_3 \in E(G^2)$ such that $v_5$ is nonadjacent to $v_1,v_2,v_4$ in $G^2$. If the pairwise distance of $v_i,v_{i+1}$ is 2 for every $i \in \{1,2,3,4\}$ (modulo 4) then it forms a hole in $G$. Thus, assume that $d_G(v_1,v_4)=1$; let $v_1 \in X$ and $v_4 \in Y$. Then, as above, $d_G(v_1,v_2)=2$ (since $v_2v_4 \notin E(G^2)$) and $d_G(v_3,v_4)=2$
 (since $v_1v_3 \notin E(G^2)$). Thus, $v_2 \in X$ and $v_3 \in Y$, and now $d_G(v_2,v_3)=1$.
Let $y_1 \in Y$ be a common neighbor of $v_1,v_2$, and let $x_3 \in X$ be a common neighbor of $v_3,v_4$. Since $G$ is hole-free, $y_1x_3 \in E$.
Then $d_G(v_3,v_5)=2$ since $v_2v_5 \notin E(G^2)$, i.e., $d_G(v_2,v_5) \ge 3$; let $x_5 \in X$ be a common neighbor of $v_3,v_5$. By the degree bound and since $deg_G(y_1)=3$, we have $y_1x_5 \notin E$. Then, $v_1,\ldots,v_5,y_1,x_3,x_5$ induce an $H_4$ (as shown in Figure~\ref{extdomH4}), which is a contradiction.

\medskip

Thus, Theorem \ref{EDchordbipgrdeg3} is shown.
\qed

\begin{corollary}\label{coro:EDchordbipgrdeg3}
For $H_4$-free chordal bipartite graphs with vertex degree at most $3$, ED is solvable in polynomial time.
\end{corollary}

Recall that ED is \NP-complete for bipartite graphs of vertex degree at most 3 \cite{BraMilNev2013} and girth at least $g$ for every fixed $g$ \cite{Nevri2014}.

\begin{theorem}\label{EDP9frbipgrdeg3}
For $P_9$-free bipartite graphs with vertex degree at most $3$, ED is solvable in polynomial time.
\end{theorem}

\noindent
{\bf Proof.}
Let $G=(V,E)$ be a $P_9$-free bipartite graph with vertex degree at most $3$. Again, by Theorem \ref{theoB-T} and by the e.d.s.\ property, when checking whether $v \in V$ is part of an e.d.s.\ $D$ of $G$, we can assume that its distance levels $N_k$, $k \ge 6$, are empty.
By Claim \ref{vforcedvnotinD} $(iii)$ and (\ref{xN3N4neighb}), we can assume that every vertex in $N_3$ has a neighbor in $N_4$. We first show:

\begin{Claim}\label{DcapN3atmost2}
$|D \cap N_3| \le 2$.
\end{Claim}

\noindent
{\em Proof.}
Suppose to the contrary that $|D \cap N_3| \ge 3$; let $y_1,y_2,y_3 \in D \cap N_3$, and let $z_i$, $i \in \{1,2,3\}$ be neighbors of $y_i$ in $N_4$ and let $x_i$, $i \in \{1,2,3\}$ be neighbors of $y_i$ in $N_2$. Clearly, by the e.d.s.\ property and by (\ref{2P3N2N3N4}), $x_1,y_1,z_1,x_2,y_2,z_2,x_3,y_3,z_3$ induce a $3P_3$ in $G$. Let $u_i \in N_1$, $i \in \{1,2,3\}$, be a common neighbor of $v$ and $x_i$. By the degree bound 3, there is no common neighbor $u \in N_1$ of $x_1,x_2$, and $x_3$.

First assume that for two of $x_1,x_2,x_3$, there is a common neighbor in $N_1$; without loss of generality, let $u_1x_1 \in E$ and $u_1x_2 \in E$ for $u_1 \in N_1$.
Then, by the degree bound 3, $u_1x_3 \notin E$, and thus, there is a distinct neighbor $u_3 \in N_1$ with $u_3x_3 \in E$.
Since $z_1,y_1,x_1,u_1,v,u_3,x_3,y_3,z_3$ do not induce a $P_9$ in $G$, we have $u_3x_1 \in E$, and
since $z_2,y_2,x_2,u_1,v,u_3,x_3,y_3,z_3$ do not induce a $P_9$ in $G$, we have $u_3x_2 \in E$, which is a contradiction to the degree bound 3.
Thus, there is no common neighbor in $N_1$ of two of the vertices $x_1,x_2,x_3$; each of $x_i$ has its private neighbor $u_i \in N_1$, $i \in \{1,2,3\}$.
But then $z_1,y_1,x_1,u_1,v,u_2,x_2,y_2,z_2$ induce a $P_9$ in $G$, which is a contradiction.
Thus $|D \cap N_3| \le 2$.
$\diamond$

\medskip

Thus, for every pair $y_1,y_2 \in N_3$, we can check whether there is an e.d.s.\ $D$ of $G$ with $v,y_1,y_2 \in D$ by reducing the graph correspondingly; let $N'_4:=N_4 \setminus (N(y_1) \cup N(y_2))$. Again, we can assume that all vertices in $N'_4$ have a neighbor in $N_5$ since otherwise, such vertices are $(v,y_1,y_2)$-forced by the assumption that $v,y_1,y_2 \in D$. By similar arguments as for Claim \ref{DcapN3atmost2}, we can show that $|D \cap N_4| \le 2$ and finally, for vertices $z_1,z_2 \in N'_4 \cap D$, $N_5 \setminus (N(z_1) \cup N(z_2)) \subset D$ is forced. Thus, Theorem \ref{EDP9frbipgrdeg3} is shown.
\qed

\medskip

By the degree bound 3, it is obvious that a bipartite graph $G$ with induced subgraph $K_{3,3}$ has no e.d.s. Moreover, for a $K_{2,3}$ with degree 3 vertices $a$ and $b$, these two vertices are excluded. What is the complexity of ED for $K_{2,3}$-free bipartite graphs with vertex degree at most 3?

\section{\NP-completeness of ED for some subclasses of bipartite graphs}

In \cite{AbrRah2018}, it is shown that ED is \NP-complete for graphs with diameter at most 3. For bipartite graphs we show:

\begin{theorem}\label{EDNPcchordalbipgr}
ED is \NP-complete for bipartite graphs with diameter at most $6$.
\end{theorem}

\noindent
{\bf Proof.} The proof is based on the transformation from the Exact Cover problem X3C to ED for bipartite graphs. Let $H=(V,{\cal E})$ with $V=\{v_1,\ldots,v_n\}$ and ${\cal E}=\{e_1,\ldots,e_m\}$ be a hypergraph with $|e_i|=3$ for all $i \in \{1,\ldots,m\}$. Let $G_H$ be the following transformation graph:

$V(G_H)=V \cup X \cup Y \cup \{z,w,u\}$ such that $X=\{x_1,\ldots,x_m\}$, $Y=\{y_1,\ldots,y_m\}$ and $V,X,Y,\{z,w,u\}$ are pairwise disjoint. The edge set $E(G_H)$ of $G_H$ consists of the following edges:
First, $v_ix_j \in E(G_H)$ whenever $v_i \in e_j$. Moreover $V$ is an independent set in $G_H$, every $y_i$ is only adjacent to $x_i$ in $G_H$, $z \join V$, and $zw,wu \in E(G_H)$.

Clearly, $G_H$ is bipartite, and it is easy to see that the diameter of $G_H$ is at most $6$.

\medskip

Next we show that $H=(V,{\cal E})$ has an exact cover if and only if $G_H$ has an e.d.s.\ $D$:

For an exact cover ${\cal E}' \subset {\cal E}$ of $H$, every $e_i \in {\cal E}'$ corresponds to vertex $x_i \in D$, and every $e_i \notin {\cal E}'$ corresponds to vertex $y_i \in D$. Moreover, $w \in D$. Thus, $D$ is an e.d.s.\ of $G_H$.

\medskip

Conversely, if $D$ is an e.d.s.\ in $G_H$ then $V \cap D = \emptyset$ since otherwise, by the e.d.s.\ property, some $y_i$ cannot be dominated. Analogously, $z \notin D$ since otherwise, $u$ cannot be dominated. Thus, $w \in D$ is forced, and now, $D \cap X$ corresponds to an exact cover of $H$, namely for $D=\{x_{i_1}, \ldots, x_{i_k}\}$, $\{N(x_{i_1}) \cap V, \ldots, N(x_{i_k}) \cap V\}$ is an exact cover of $H$.
\qed

\medskip

This proof can be easily extended for showing that ED is \NP-complete for $C_4$-free bipartite graphs (and more generally, for $(C_4,C_6,\ldots,C_{2k})$-free bipartite graphs for every fixed $k \ge 3$): As a first step, every edge $x_iv_j \in E(G_H)$ has to be replaced by a $P_5$ $P(x_i,v_j)$ with end-vertices $x_i$ and $v_j$ and private internal vertices. Let $P(x_i,v_j)=(x_i,a,b,c,v_j)$. If $x_i \in D$ then $c \in D$ is forced, and if $x_i \notin D$ then $b \in D$ is forced.
Clearly, the replacement of $G_H$ is $C_4$-free, and iteratively, the replacement is $(C_4,C_6,\ldots,C_{2k})$-free).

The result of Nevries \cite{Nevri2014} that ED is \NP-complete for planar bipartite graphs with degree 3 and girth at least $g$ has been mentioned already but the special case of $(C_4,C_6,\ldots,C_{2k})$-free bipartite graphs mentioned above is much easier to prove (clearly, the iterative replacement of $G_H$ is not planar bipartite and does not have vertex degree 3 but has girth at least $g$).

\section{Conclusion}

{\bf Open problems}: What is the complexity of ED for $P_k$-free bipartite graphs, $k \ge 8$, for $S_{1,3,3}$-free bipartite graphs, for $S_{1,1,5}$-free bipartite graphs, and in general for $S_{2,2,k}$-free bipartite graphs for $k \ge 5$, and for chordal bipartite graphs with vertex degree at most $3$?

\medskip

\noindent
{\bf Acknowledgment.} The second author would like to witness that he just tries to pray a lot and is not able to do anything without that - ad laudem Domini.

\begin{footnotesize}

\end{footnotesize}

\end{document}